\documentclass[11pt]{article} 
\usepackage{hyperref} 
\pdfoutput=1 
 
\begin{document} 
 
\title{Feeding Currents Generated by Upside Down Jellyfish} 
 
\author{Terry Rodriguez, Christina Hamlet, Megan Gyoerkoe, and Laura Miller \\ 
\\\vspace{6pt} Department of Mathematics, University of North Carolina at Chapel Hill, \\ Chapel Hill, NC 27599, USA} 
 
\maketitle 
 
 
\begin{abstract} 
We present fluid dynamics videos of the pulsing dynamics and the resulting fluid flow generated by the upside down jellyfish, \textit{Cassiopea spp}. Medusae of this genus are unusual in that they typically rest upside down on the ocean floor and pulse their bells to generate feeding currents, only swimming when significantly disturbed. The pulsing kinematics and fluid flow around these upside-down jellyfish is investigated using a combination of videography, flow visualization, and numerical simulation. Significant mixing occurs around and directly above the oral arms and secondary mouths. Numerical simulations using the immersed boundary method with a porous layer representing the oral arms agree with the experimental results. The simulations also suggest that the presence of porous oral arms induce net horizontal flow towards the bell. Coherent vortex rings are not seen in the wake above the jellyfish, but starting and stopping vortices are observed before breaking up as they pass through the elaborate oral arms (if extended). 
\end{abstract} 
 
 
\section{Introduction} 
 The jellyfish has been the subject of numerous mathematical and physical studies ranging from the discovery of reentry phenomenon in electrophysiology to the development of axisymmetric methods for solving fluid-structure interaction problems. A wide variety of experimental, theoretical, and numerical studies have been performed to understand how jellyfish propel themselves through the water using bell pulsations. More recently, experimental and numerical studies have been performed to understand how jellyfish with relatively simple morphologies use bell pulsations to locomote and to feed. In this work we capitalize upon the unique properties of the upside down jellyfish, \textit{Cassiopea spp.}, to understand the dynamics of feeding currents generated through complex oral arm structures and uncoupled from locomotion. The upside down jellyfish is ideal for such work since they typically rest on the sandy bottom of the ocean with their oral arms facing upwards and towards the sun. Bell pulsations are used primarily to drive flow through an elaborate array of oral arms.
 
The fluid dynamics videos are available as a \href{http://frg.unc.edu/movies/Upsidedownjelly.mp4}{larger file size} and as a \href{http://frg.unc.edu/movies/Upsidedownjelly_small.mp4}{smaller file size}. Flow visualization is used to reveal the generation of starting and stopping vortices that then break up into smaller structures as they advect through the oral arms. Numerical simulations show similar effects when the oral arms are added as a porous layer above the pulsing bell.

\section{Methods}
\textit{Cassiopea spp.} medusae were ordered from Carolina Biological Supply and Gulf Specimen Marine Supply and were housed in 29 gallon aquaria. Flow visualization was performed using gravity fed fluorescein and rhodamine dyed. The dye was injected into the sand below the jellyfish and was slowly pulled to the surface through the pulsation of the jellyfish bells. Black lights were used to illuminate the dyes against the black background. The immersed boundary method was then used to solve the fluid-structure interaction problem and explore how changes in morphology and pulsing dynamics alter the resulting fluid flow~\cite{Peskin}. The oral arms were represented as a porous layer using the method described by Kim and Peskin~\cite{Kim} and Stockie~\cite{Stockie}. Visualizations of the computational results were performed using $DataTank^{TM}$. The experimental and computational components of this work were conducted at the University of North Carolina at Chapel Hill, in the facilities of the Mathematical Physiology Laboratory.

\end{document}